\documentclass [12pt]{article}

\begin{document}

\title{\bf Ultra-Relativistic Expansion of Ideal Fluid with Linear
Equation of State }

\author{V.I.Zhdanov, M.S.Borshch \\
\small Kyiv National Taras Shevchenko University\\
}

\date{}

\maketitle

\begin {abstract}

We study solutions of the relativistic hydrodynamical equations,
which describe spherical or cylindrical expansion of ideal fluid.
We derived approximate solutions involving two arbitrary
functions, which describe asymptotic behavior of expanding
fireballs in ultra-relativistic limit. In case of a linear
equation of state $p(\varepsilon )= \kappa \varepsilon - c_1 $, $
(0<\kappa<1)$ we show that the solution may be represented in form
of an asymptotic series in negative powers of radial variable;
recurrence relations for the coefficients are obtained. This
representation is effective, if $\kappa > 1/(2J+1)$ ($J=2$ for
spherical expansion and $J=1$ for the cylindrical one); in this
case the approximate solutions have a wave-like behavior.

\end{abstract}


\section{Introduction }

The problem of the relativistic ideal fluid expansion arise, e.g.,
in the theory of gamma-ray bursts, that represent the most
powerful explosions in the Universe \cite{Mez}. The other
applications are due to hydrodynamical theory of multiparticle
production \cite{Ochelk}. Here the relativistic ideal fluid model
is utilized to describe the behavior of relativistic fireballs.
Detailed investigation of this problem involves complicated 3-D
numerical simulations. Nevertheless, considerable information may
be obtained from analytical studies, which  deal typically with
one-dimensional flows having spherical, cylindrical and plane
symmetry \cite{Sibg}, \cite{Ochelk}. One of such well-known
results is due to Blandford {\&} McKee \cite{BlMcKee}, who studied
self-similar ultra-relativistic gas expansion following a strong
shock. Blandford {\&} McKee solution is widely used for
interpretation of the gamma-ray bursts and their afterglows
\cite{Mez}. Certain drawbacks of this solution have been pointed
out in \cite{Gruzinov}.

It should be noted, that investigation of ultra-relativistic
asymptotics of ideal flows is rather specific; it requires some
care to provide approximation correctness. For example, if we
leave only the main terms in the hydrodynamical equations in
ultra-relativistic limit, these become degenerate (see Sect. 2).
To be sure about approximation involved, one must study the higher
order corrections in the equations. Evidently, it is desirable to
have a sufficiently general form of the solutions that contain
necessary number of arbitrary functions. This is important, e.g.,
in order to describe models of expanding fluid with different
radial density profiles.

In this connection we propose an approximation method to construct
general solutions that describe ultra-relativistic radial
expansion of ideal fluid with necessary degree of accuracy. Most
detailed result is obtained in case of a linear equation of state
(EOS); in this case we represent the solution in form of
asymptotic series (Sect. 3), which is workable at least for a
sufficiently stiff EOS. These results are compared in Sect. 4 with
a self-similar solutions.

\section{Main equations }

The equations of motion of ideal relativistic fluid follow from
the conservation laws (see, e.g., \cite{Ochelk}\cite{Sibg})

\[
\partial _\nu T^{\mu \nu } = 0,
\]

\noindent for the energy-momentum tensor $T^{\mu \nu
}=(p+\varepsilon)u^{\mu }u^{\nu }-pg^{\mu \nu }$;  $u^{\nu }$ is
the fluid four -velocity, $g^{\mu \nu }=diag(1,-1,-1,-1)$, $p$ is
the pressure, $\varepsilon $ is the invariant energy density, the
light speed $c=1$. We confine ourselves to EOS $p=p(\varepsilon)$.
Some kind of the continuity equation must be addedIn in a more
general case of two-parametric EOS.

In case of cylindrical ($J=1$) and spherical ($J=2$) symmetry of
the flow we deal with only two independent variables (time $t$ and
radial variable $r$). The equations take on the form \cite{Sibg},
\cite{Ochelk}:

\begin{equation}
\label{eq1}
  \frac{\partial }{\partial t}\left[
{(u^t)^2\varepsilon + (u^r)^2p} \right] + \frac{\partial
}{\partial r}\left[ {u^tu^r(p + \varepsilon )} \right] = -
\frac{J(p + \varepsilon )}{r}u^tu^r
,
\end{equation}

\begin{equation}
\label{eq2}
 \frac{\partial }{\partial t}\left[ {u^tu^r(p +
\varepsilon )} \right] + \frac{\partial }{\partial r}\left[
{(u^t)^2p + (u^r)^2\varepsilon } \right] = - \frac{J(p +
\varepsilon )}{r}(u^r)^2 ,
\end{equation}

where $u^r$ is the radial component of the four-velocity, the
other spatial four-velocity components are equal to zero.

In ultra-relativistic limit $u^r \approx u^t >> 1 $. But if we
leave in Eqs.(\ref{eq1}) and (\ref{eq2}) only the higher order
terms in $u^{r}$, then both from (\ref{eq1}) and (\ref{eq2}) we
have the same equation

\[
\frac{\partial }{\partial t}\left[ {(u^t)^2(\varepsilon + p)}
\right] + \frac{\partial }{\partial r}\left[ { (u^t)^2(\varepsilon
+ p)} \right] = - \frac{J}{r}\left[ {(u^t)^2(\varepsilon + p)}
\right],
\]

\noindent that is we have certain degeneracy. We can determine
$(u^t)^2(\varepsilon + p)$ from this equation, but we cannot
determine the velocity and the energy density separately. This
shows that we cannot neglect the terms, looking at first sight as
less important, without more detailed consideration.

\section{First approximation}

Further we use the substitution $u^t = u + 1/(4u)$, $u^r = u -
1/(4u)$. Then Eqs. (\ref{eq1}) and (\ref{eq2}) yield:

\[
 \frac{\partial }{\partial t}\left[ {\left( {u^2 +
\frac{1}{16u^2}} \right)\left( {\varepsilon + p} \right) +
\frac{\varepsilon - p}{2}} \right] +\frac{\partial }{\partial
r}\left[ {\left( {u^2 - \frac{1}{16u^2}} \right)\left(
{\varepsilon + p} \right)} \right]=
\]
\begin{equation}
\label{eq1u} =  - \frac{J\left( {p + \varepsilon }
\right)}{r}\left( {u^2 - \frac{1}{16u^2}} \right),
\end {equation}

\[
 \frac{\partial }{\partial t} \left[ {\left(
{u^2-\frac{1}{16u^2}}
 \right)\left( {\varepsilon + p}\right)}\right]
+ \frac{\partial }{\partial r}\left[ {\left( {u^2 +
\frac{1}{16u^2}} \right)\left( {p + \varepsilon } \right) -
\frac{\varepsilon - p}{2}} \right]=
\]
\begin{equation}
\label{eq2u}
 = - \frac{J\left( {p + \varepsilon } \right)}{r}\left( {u^2 +
\frac{1}{16u^2} - \frac{1}{2}} \right).
\end{equation}

Consider the sum and difference of Eqs. (\ref{eq1u}) and
(\ref{eq2u}) in new variables $\alpha =t-r$, $\beta =t+r$ and . We
have

\begin{equation}
\label{eq1Nvar} \frac{\partial }{\partial \beta }\left(
{\varepsilon - p} \right) + \frac{\partial }{\partial \alpha
}\left( {\frac{\varepsilon + p}{4u^2}} \right) = - \frac{J\left(
{\varepsilon + p} \right)}{\beta - \alpha }\left( {1 -
\frac{1}{4u^2}} \right)
\end{equation}

\begin{equation}
\label{eq2Nvar} \frac{\partial }{\partial \beta }\left[ {u^2\left(
{\varepsilon + p} \right)} \right] + \frac{\partial }{\partial
\alpha }\left( {\frac{\varepsilon - p}{4}} \right) = -
\frac{J\left( {\varepsilon + p} \right)}{\beta - \alpha }\left(
{u^2 - \frac{1}{4}} \right)
\end{equation}

These are still exact equations.

In the ultra-relativistic approximation $u>>1$ we leave the terms
$\sim u^2$ and zero orders in this value. Neglecting the terms
containing $\sim 1/u^2$, we have from (\ref{eq1Nvar})

\begin{equation}
\label{eq5} \frac{\partial }{\partial \beta }\left( {\varepsilon -
p} \right) \approx - \frac{J\left( {\varepsilon + p}
\right)}{\beta - \alpha }.
\end{equation}

\noindent This equation can be solved, if we know  the state
equation $p=p(\varepsilon)$. In this case Eq. (\ref{eq5}) yields

\begin{equation}
\label{eqaa} \left( 1 - dp/d\varepsilon \right)\frac{\partial
\varepsilon }{\partial \beta } \approx - \frac{J\left[
{\varepsilon + p\left( \varepsilon \right)} \right]}{\beta -
\alpha } .
\end{equation}

Let by definition
\[
  \Phi ( \varepsilon ) = exp \left[{ \int {\frac{1 - dp/d \varepsilon
}{\varepsilon + p\left( \varepsilon \right)}d\varepsilon }
}\right].
\]

 Then the solution of Eq.(\ref{eqaa}) may be
represented as

\begin{equation}
\label{solappreps} \varepsilon \approx \Phi ^{ - 1}\left\{
{\frac{f\left( \alpha \right)}{\left( {\beta - \alpha }
\right)^J}} \right\},
\end{equation}

\noindent where $f(\alpha)$ is an arbitrary function. The validity
of this approximate solution depends on smallness of the terms in
Eq. (\ref{eq1Nvar}) that have been neglected, i.e. the
approximation is valid if

\[
\left|{ \int { d \beta (\beta-\alpha)^J \frac{\partial }{\partial
\alpha } \left( { \frac{\varepsilon + p}{4u^2(\beta-\alpha)^J }
}\right)}}\right|<<|f(\alpha)|.
\]

As we shall see in the next section, this condition is satisfied,
e.g., in case of a linear EOS $p(\varepsilon )= \kappa \varepsilon
- c_1 $, if $1> \kappa > 1/(2J+1)$.

Eq.(\ref{eq2Nvar}) can be written as

\[
(\beta-\alpha)^{-J}\frac{\partial }{\partial \beta }\left[ {u^2
(\beta-\alpha)^J \left( {\varepsilon + p} \right)} \right]
 = \frac{J(\varepsilon + p)}{4(\beta-\alpha)}-
\frac{\partial}{\partial\alpha}\frac{\varepsilon-p}{4}.
\]

Using $\varepsilon$ from (\ref{solappreps}) we obtain $u$ from

\[
u^2 (p + \varepsilon)( \beta - \alpha )^{J } =
\]

\begin{equation}
\label{solappu} = g( \alpha ) +  \int {d\beta
\frac{(\beta-\alpha)^J}{4} \left[{ \frac{J(\varepsilon +
p)}{(\beta-\alpha)}-\frac{\partial}{\partial\alpha}(\varepsilon-p)}\right]},
\end{equation}

\noindent where $g(\alpha)$ is an arbitrary function. This enables
us to find $\varepsilon$ and $u$ separately.

 If the integral in the right-hand side of  (\ref{solappu}) converges for
$\beta \to \infty$ and $\alpha$ remains bounded, then for
sufficiently large $\beta \sim t$ we obtain that $ u^2 (
\varepsilon+p)( \beta - \alpha )^{J }$ is a function only of
$\alpha$. We shall see below that this is just the case of the
linear EOS with $\kappa > 1/(2J+1)$. In this case we have that the
energy profile of the ultra-relativistic expanding shell $dE/d
\alpha \approx (\varepsilon+p) u^2 r^J $ remains almost constant
for large values of $t$.

\section{Higher orders of asymptotic series for the solutions}

Consider the higher orders of approximation in case of a linear
equation of state $p(\varepsilon) = \kappa \varepsilon - c_1 $,
where $\kappa=c_0^2<1$, $c_0$ is a speed of sound. We are looking
for the asymptotic representation of the solution, which is
workable for bounded $\alpha= t-r$ and $ t \to \infty $; this
means also $\beta \sim r \to \infty$.

We put

\[
\varepsilon = \frac{\varphi ( \alpha ,\beta )}{(\beta -
\alpha)^a}+c_1/(1+\kappa),
 \quad
 a = \frac{J\left( {1 + \kappa } \right)}{1 -
\kappa },
\]
\[
u^2(\varepsilon + p) = \frac{\psi (\alpha ,\beta )}{(\beta -
\alpha )^J}.
\]

This substitution reduces the system (\ref{eq1Nvar}),
(\ref{eq2Nvar}) to the form

\[
\frac{\partial \varphi }{\partial \beta } + \frac{\left( {1 +
\kappa } \right)^2}{4\left( {1 - \kappa } \right)}\frac{1}{\left(
{\beta - \alpha } \right)^{a - J}}\frac{\partial }{\partial \alpha
}\left( {\frac{\varphi ^2}{\psi }} \right) =
\]

\begin{equation}
\label{eq7}
 = - \frac{J\kappa \left( {1 + \kappa } \right)^2}{\left( {1 - \kappa }
\right)^2}\frac{1}{\left( {\beta - \alpha } \right)^{a - J +
1}}\frac{\varphi ^2}{\psi }
\end{equation}

\begin{equation}
\label{eq8} \frac{\partial \psi }{\partial \beta } + \frac{1 -
\kappa }{4}\frac{1}{\left( {\beta - \alpha } \right)^{a -
J}}\frac{\partial \varphi }{\partial \alpha } = 0
\end{equation}

We shall look for solution of exact equations
(\ref{eq7}),(\ref{eq8}) in the form of asymptotic series
($m,n=0,1,2,...$)

\begin{equation}
\label{rawfi} \varphi \left( {\alpha ,\beta } \right) =
\sum\limits_{n, m} {\frac{\varphi _{n, m} \left( \alpha
\right)}{\left( {\beta - \alpha } \right)^{\gamma n + m}}},
\end{equation}

\begin{equation}
\label{rawpsi} \psi \left( {\alpha ,\beta } \right) =
\sum\limits_{n,m} {\frac{\psi _{n,m} \left( \alpha \right)}{\left(
{\beta - \alpha } \right)^{\gamma n + m}}} ,
\end{equation}
\noindent where $\gamma = a-J-1$.

If we are looking for an asymptotic solution for $r \to \infty $,
this representation is effective for $\gamma > 0$, that is for
$\kappa > 1 / (2J+1)$.

We also introduce

\[
\chi \left( {\alpha ,\beta } \right) = \frac{\varphi ^2}{\psi } =
\sum\limits_{n,m} {\frac{\chi _{n,m} \left( \alpha \right)}{\left( {\beta -
\alpha } \right)^{\gamma n + m}}} ;
\]

\noindent the coefficients of these series may be expressed be
means of $\varphi_{n,m }$, $\psi_{n,m}$ . Moreover, $\chi _{n,m}$
depends only upon  $\varphi_{n',m' }$, $\psi_{n',m'}$ with $n'\leq
n$, $m' \leq m$.

We now substitute the above representations for $\varphi$ , $\psi
$, $\chi $ into Eqs.(\ref{eq7}), (\ref{eq8}) to find recurrence
relations for $\varphi _{n,m }$, $\psi_{n,m}$.

For $n \ge 1, m \ge 1$ we have

\[
  \left( {\gamma n + m} \right)\varphi _{n,m}=
\]
\begin{equation}
\label{eq9} =\frac{\left( {1 + \kappa } \right)^2}{4\left( {1 -
\kappa } \right)} \left\{ { \frac{d \chi _{n - 1,m} }{d \alpha } +
\left[ { \gamma (n - 1) + m - 1 + \frac {4J \kappa}{(1-\kappa) } }
\right] \chi_{n -1,m - 1} } \right\},
\end{equation}

\[
  \left( {\gamma n + m} \right)\psi _{n,m}=
\]
\begin{equation}
\label{eq10} = \frac{1 - \kappa }{4}\frac{d \varphi _{n - 1,m} }{d
\alpha } + \frac{1 - \kappa }{4}\left[ {\gamma \left( {n - 1}
\right) + m - 1} \right]\varphi _{n - 1,m - 1}.
\end{equation}

For $n=0, m=1,2,...$

\begin{equation}
\label{eq9a} \varphi_{0,m}=\psi_{0.m}=0,
\end{equation}

whence also $\chi_{0,m}=0$.

For $m=0, n=1,2,...$

\begin{equation}
\label{eq10a} n \gamma \varphi _{n,m}=\frac{\left( {1 + \kappa }
\right)^2}{4\left( {1 - \kappa } \right)}\frac{d \chi _{n - 1,0}
}{d \alpha }
\end{equation}
and
\begin{equation}
\label{eq10b}
   \gamma n  \psi _{n,0}=\frac{1 - \kappa }{4}\frac{d \varphi _{n - 1,0} }{d
\alpha }
\end{equation}

Using (\ref{eq9}),(\ref{eq10}), (\ref{eq9a}), (\ref{eq10a}) we can
express all the coefficients of the series
(\ref{rawfi}),(\ref{rawpsi}) by means of $\psi_{00}(\alpha)$ and
$\varphi_{00}(\alpha)$ ($\psi_{00}\ne 0$).

Here we write expressions for some lowest order coefficients

\[
\varphi _{1,0} = \frac{\left( {1 + \kappa } \right)^2}{4 \gamma
\left( {1 - \kappa } \right) }\frac{d \chi _{0,0} }{d \alpha },
\quad \psi _{1,0} = \frac{1 - \kappa }{4\gamma }\frac{d \varphi
_{0,0} }{d \alpha },
\]
\[
\varphi _{2,0} = \frac{(1 + \kappa )^2}{8 \gamma (1 - \kappa)}
\frac{d \chi _{1,0} }{d \alpha }, \quad \chi_{1,0}=2
\frac{\varphi_{1,0}\varphi_{0,0}}{\psi_{0,0}}-\frac{\psi_{1,0}\varphi_{0,0}^2}{\psi_{0,0}^2},
\]
\[
\psi _{2,0} = \frac{\left( {1 + \kappa } \right)^2}{2\gamma
^2}\frac{d ^2\chi _{0,0} }{d \alpha ^2},\quad \varphi _{1,1} =
\frac{J\kappa \left( {1 + \kappa } \right)^2}{\left( {1 - \kappa }
\right)^2\left( {\gamma + 1} \right)}\chi _{0,0},
\]

$ \quad \psi _{1,m} = 0, m = 1,2,..., $

\section{Self -similar solutions }

The representation of solution in the previous section allows us
to obtain the coefficients of the series
(\ref{rawfi}),(\ref{rawpsi}) up to any given order. Therefore, if
$\kappa > 1/(2J+1)$, we may calculate asymptotic representation of
the solution with any desired accuracy for $r \to \infty$.
Nevertheless, this does not mean that series (\ref{rawfi}) and
(\ref{rawpsi}) are convergent. Therefore, it would be useful to
check the results of the previous section using the self-similar
solutions, when we deal with ordinary differential equations.

We choose the similarity variable as $\xi=r/t $. In case of the
EOS $p=\kappa \varepsilon -c_1/(1+\kappa)$ the system
(\ref{eq1})-(\ref{eq2}) leads to the ordinary differential
equations

\begin{equation}
\label{eq1self} \xi \frac{ dv}{d\xi} =J\kappa (1-v^2) \frac{\mu\xi
(1- v^2)+v(1-v\xi)} {[(v-\xi)^2-\kappa(1-v\xi)^2]},
\end{equation}

\begin{equation}
\label{eq2self} \xi \frac{d\sigma }{d\xi} = J(1+\kappa)
\frac{\mu\kappa+[1+(1-\kappa)\mu]v\xi-(1+\mu)v^2}
{[(v-\xi)^2-\kappa(1-v\xi)^2]},
\end{equation}

\noindent where $v=u^r/u^t=(4u^2-1)/(4u^2+1)$,  $\sigma =ln\rho
,\quad \varepsilon = r^\nu \rho +c_1/(1+\kappa)$,  $\mu = \nu/[J(1
+ \kappa ) ]$.

A qualitative investigation of self-similar relativistic fluid
flows with spherical and cylindrical symmetry can be found in
\cite{Sibg}. Here we confine ourselves to asymptotical solutions
of the Eqs. (\ref{eq1self}),(\ref{eq2self}) near the boundary of
expanding fluid, where $u>>1$, $v\to 1$. If the fluid expands to
vacuum, its boundary moves with the light speed (correspondingly,
$\xi=1$). Using the results of qualitative invesigation of
\cite{Sibg} in the phase plane $\{v,\xi\}$, it is easy to classify
possible asymptotic solutions of Eqs.(\ref{eq1self}),
(\ref{eq2self}) near this boundary for $\xi \to 1$ and to obtain
their asymptotics.

In case of  $\kappa > 1/(2J+1)$ we have

\begin{equation}
\label{eq18}
 u^2 \approx A_1 (1-\xi)^{p_1}, \quad p_1 = -\frac{2\kappa J}{1 - \kappa },
\end {equation}

\begin{equation}
\label{asymp1}
 \varepsilon \approx A_2 r^\nu(1-\xi)^{p_2}, \quad p_2 =
\nu+\frac{J(1+\kappa)}{1-\kappa} \quad ,
\end {equation}

\noindent $A_1,A_2$ being arbitrary constants.

This corresponds to the results of the previous section, where one
should assign

\[
\varphi_{00}(\alpha)\sim \alpha^{p_2}, \quad \psi_{00}(\alpha)\sim
\alpha^{\nu+J}.
\]

If $\kappa < 1/(2J+1)$, this type of solution is absent.

 On the other hand, there is also a solution with asymptotics

\begin{equation}
\label{asymp2} u^2=A_{\pm} (1-\xi)^{-1},\quad
A_{\pm}=(1-\kappa)^{-1}(B\pm \sqrt{D}),
\end{equation}

where
\[
 B=1+\kappa+J\kappa(1+2\mu), \quad
D=B^2+(1-\kappa)[(1+2J)\kappa-1].
\]

 For $\kappa > 1/(2J+1)$ it is easy to see that
$A_- <0$, so only $A_+>0$ survives. For $\kappa < 1/(2J+1)$ both
$A_{-}$ and $A_{+}$ are positive. Depending upon $\nu$ and
$\kappa$, Eq.(\ref{asymp2}) may represent either a special
solution or an infinite family of solutions. Some values of the
parameters may lead to $\varepsilon \to \infty$  for ${\xi \to 1}$
and they must be matched to another solutions through the shock
wave; the other parameters correspond to solutions that can be
matched to vacuum ($\varepsilon \to 0$). This asymptotics is not
well described by a finite number of terms in the representation
(\ref{rawfi}), (\ref{rawpsi}) of the previous section.

\section {Discussion}

We studied  non-stationary ultra-relativistic hydrodynamical flows
having spherical or cylindrical symmetry in case of the equation
of state $p = p(\varepsilon )$. The results may be applicable also
in case of a more general EOS if the fluid motion is isentropic.
The solutions derived  in Sect. 4 (as well as the first order
approximations of Sect. 3) contain two arbitrary functions,
therefore they represent some kind of a general solution of the
equations of the fluid motion. However, consideration of the
section 5 shows that  they do not describe solutions of all
possible types equally well in the ultra-relativistic limit.
Formally we may use some of the above results in case of arbitrary
$\kappa \in (0,1)$, $p(\varepsilon)=\kappa\varepsilon-c_1$, in any
ultra-relativistic case, but in certain limited domain of spatial
variables, where the approximations work.

The approximations  work in the best way in case of the "stiff"
equation of state: $\kappa > 1 / 5$ in case of a spherical
expansion ( this involves the important case $\kappa = 1 /3$), and
$\kappa
> 1 / 3$ in case of a cylindrical one. In these cases the
representations of Sect. 3 are most effective for $t-r << t \to
\infty $. Corresponding solutions exhibit the wave-like behavior
irrespectively of energy profile of the expanding shell. The
energy of the spherical/cylindrical shell, which expands almost
with the light speed, is approximately conserved. This does not
mean, however, that this shell will contain all the energy of the
expanding fluid. It would be interesting to study, what is the
part of the total energy carried by this expanding wave-like
shell, and to relate the shell profile with the initial
conditions; this probably cannot be done without numerical
investigation. On the other hand, in case of  small values of
$\kappa$ we may expect that the energy of the fluid is spread over
all the values of $r \sim t$. The values of the speed of sound
$c_0^2=1/(2J+1)$ may be considered as some bifurcation points,
which separate different regimes of ultra-relativistic flow.

The method of Section 3 allows  to derive formally the solutions
up to any order of approximation. However, we do not make here any
statement about the convergence of the series, taking in mind that
calculation of high orders involves higher order derivatives.
Nevertheless, our results show that, in spite of some degeneracy
in the main order of hydrodynamical equations in
ultra-relativistic limit, we do not meet any problems like small
denominators in the higher orders and in this sense we may be sure
about the asymptotic properties of the solutions.

\end{document}